\documentclass[pdflatex,sn-mathphys-num,iicol]{sn-jnl}

\usepackage{graphicx}%
\usepackage{multirow}%
\usepackage{amsmath,amssymb,amsfonts}%
\usepackage{amsthm}%
\usepackage{mathrsfs}%
\usepackage[title]{appendix}%
\usepackage{xcolor}%
\usepackage{textcomp}%
\usepackage{manyfoot}%
\usepackage{booktabs}%
\usepackage{algorithm}%
\usepackage{algorithmicx}%
\usepackage{algpseudocode}%
\usepackage{listings}%
\usepackage{upgreek}
\usepackage{caption}

\begin{document}

\title[Article Title]{Multiscale carrier-envelope phase characterization of 2-µm~pulses delivered by a 200-kHz optical parametric amplifier}

\author*[1]{\fnm{Katrin} \sur{Meier}}\email{katrin.meier@uol.de}
\author[1]{\fnm{Arvid} \sur{Kl\"osgen}}
\author[1]{\fnm{Kerstin} \sur{Harland}}
\author[1]{\fnm{Lina} \sur{Hansen}}
\author[1]{\fnm{Lars} \sur{Barnekow}}
\author[2]{\fnm{Chen} \sur{Guo}}
\author[2]{\fnm{Caroline} \sur{Juliano}}
\author[2]{\fnm{Jin} \sur{Niu}}
\author[2]{\fnm{Cord L.} \sur{Arnold}}
\author*[1]{\fnm{Jan} \sur{Vogelsang}}\email{jan.vogelsang@uol.de}

\affil[1]{\orgdiv{Institute of Physics}, \orgname{Carl von Ossietzky Universität Oldenburg}, \orgaddress{\country{Germany}}}
\affil[2]{\orgdiv{Department of Physics}, \orgname{Lund University}, \orgaddress{\country{Sweden}}}

\abstract{Light fields with a central wavelength of 2\:µm are very well suited for strong-field-driven charge carrier control:
Their photon energy lies far below the band gap of many materials, while their oscillation period remains significantly shorter than the coherence time of charge carrier oscillations. 
The resulting potential for field-driven charge carrier control is contingent on the reproducibility of the field structure of such ultrashort laser pulses.
Here, we present a compact 200-kHz laser system that delivers ultrashort pulses with a duration of less than 20\:fs in the spectral range around 2\:µm and with a pulse energy of 25\:µJ.
The electric field structure of the 2-µm pulses is characterized in detail. In particular, the carrier-envelope phase (CEP) is measured over a wide range of timescales, from microseconds to hours. Passive stabilization due to difference frequency generation results in a root mean square value of carrier-envelope phase noise of less than 70\:mrad over all measured time scales.
The applicability of the pulses is demonstrated by measuring CEP-dependent high-order harmonic spectra with energies of up to 160\:eV.}


\maketitle

\section{Introduction}\label{sec1}
When light impinges on an atom, the electric field of the light wave interacts with electrons on a sub-femtosecond time scale \cite{Goulielmakis.2007}. 
This ultrafast interaction determines the subsequent progression of physical processes, particularly exerting a significant impact on macroscopic observables that are highly relevant for technological applications, such as all-optical communication and signal processing \cite{Rossi.2002, Wada.2004, Langer.2018}.
In nanostructures, it has been demonstrated how the instantaneous field distribution influences the photoemission of charge carriers in space and time and how a strong light field, in combination with the local field distribution around nanostructures, can be used for spatiotemporal control of electron trajectories \cite{Kruger.2011, Piglosiewicz_2014, Sumann.2015}.
The quantum control of electric currents in semiconductors, potentially before incoherent scattering processes disturb the stable phase relationship between light and matter, is of immediate technological relevance \cite{Baykusheva.2026, Langer.2020, Boolakee.2022, Schmuck.2026}.\par
Real space control of charges in the heterogeneous environment of a device is challenging due to the short quiver lengths of charge carriers that are, at best, in the single-digit nanometer range \cite{Herink_2012}.
More immediate access is available in momentum space, where light-field-driven valleytronics and Floquet engineering, for example, open up new control possibilities of bound charge carriers \cite{Langer.2018, Schaibley.2016, Choi.2025, Merboldt.2025}.
For such strong-field momentum space control, it is essential to use low photon energies to avoid or reduce photoelectron emission \cite{Baykusheva.2026, Herink_2012}.
Wavelengths in the range of 2\:µm are particularly well-suited to this application, as their photon energy of just over 0.6\:eV is significantly below the band gaps of many materials. 
Simultaneously, the oscillation period of approximately 7\:fs remains within or below the incoherent scattering times in a wide range of materials \cite{Lin.1987, Chen.2020}, whereas longer oscillation periods restrict applications, for example, to topologically protected states \cite{Ito.2023}.\par
The distributions of excited charge carriers can be investigated by detecting the photoelectrons emitted by a probe pulse with a higher photon energy \cite{Mikkelsen.2009}. To preserve the momentum information content of the photoelectrons, high photoemission currents and thus space charge effects need to be avoided \cite{Oloff_2016}.
To obtain sufficient statistics with moderate numbers of photoelectrons per shot, this requires laser pulse repetition rates in the 100\:kHz to low MHz range \cite{Chew_2014}. 
Yb-based pump lasers have enabled this development by providing high-energy femtosecond light pulses at corresponding repetition rates \cite{Rothhardt.2017}. The high repetition rate poses a characterization problem, as it makes it particularly challenging to accurately determine the stability of the electric field, and especially the carrier-envelope phase (CEP), from pulse to pulse.
Considering measurement times of at least seconds for typical experimental snapshots involving electrons, fluctuations on a shorter time scale reduce the contrast of the desired field-driven dynamics \cite{Vogelsang.2014}. Similarly, longer time scales from minutes to hours are relevant for parameter scans of multiple dimensions, and only a controlled electric field allows for a detailed analysis of the data. So far, measurements of the shot-to-shot CEP fluctuations of laser systems delivering pulses around 2\:µm with a high repetition rate are not available, as previous works averaged over the shorter time scales \cite{Homann.2012, NeuhausLaser, LundLaser, Kurihara.2023, Hergert.2024, Ritzkowsky.2026}.
Consequently, predictions for the visibility of field-driven phenomena are not possible, and complex experiments are started on the assumption that short-term fluctuations are not too severe.\par
Here we present a laser system that delivers few-cycle light pulses with a central wavelength of 2\:µm and 25\:µJ pulse energy at a repetition rate of 200\:kHz for strong-field experiments and, in particular, high-order harmonic generation (HHG).
We characterize the CEP temporally on 10 orders of magnitude, from microseconds to hours, and find residual CEP fluctuations with a root mean square (RMS) value of less than 70\:mrad at all times.
We verify the capabilities of the highly CEP-stable system by driving HHG achieving a cut-off energy of 160\:eV.
Both the CEP stability and control are demonstrated by a strong $\uppi$-periodic modulation of the extreme ultraviolet spectrum.

\section{\label{sec2:level1}Setup}

\begin{figure*}
\centering
\includegraphics[width=0.9\linewidth]{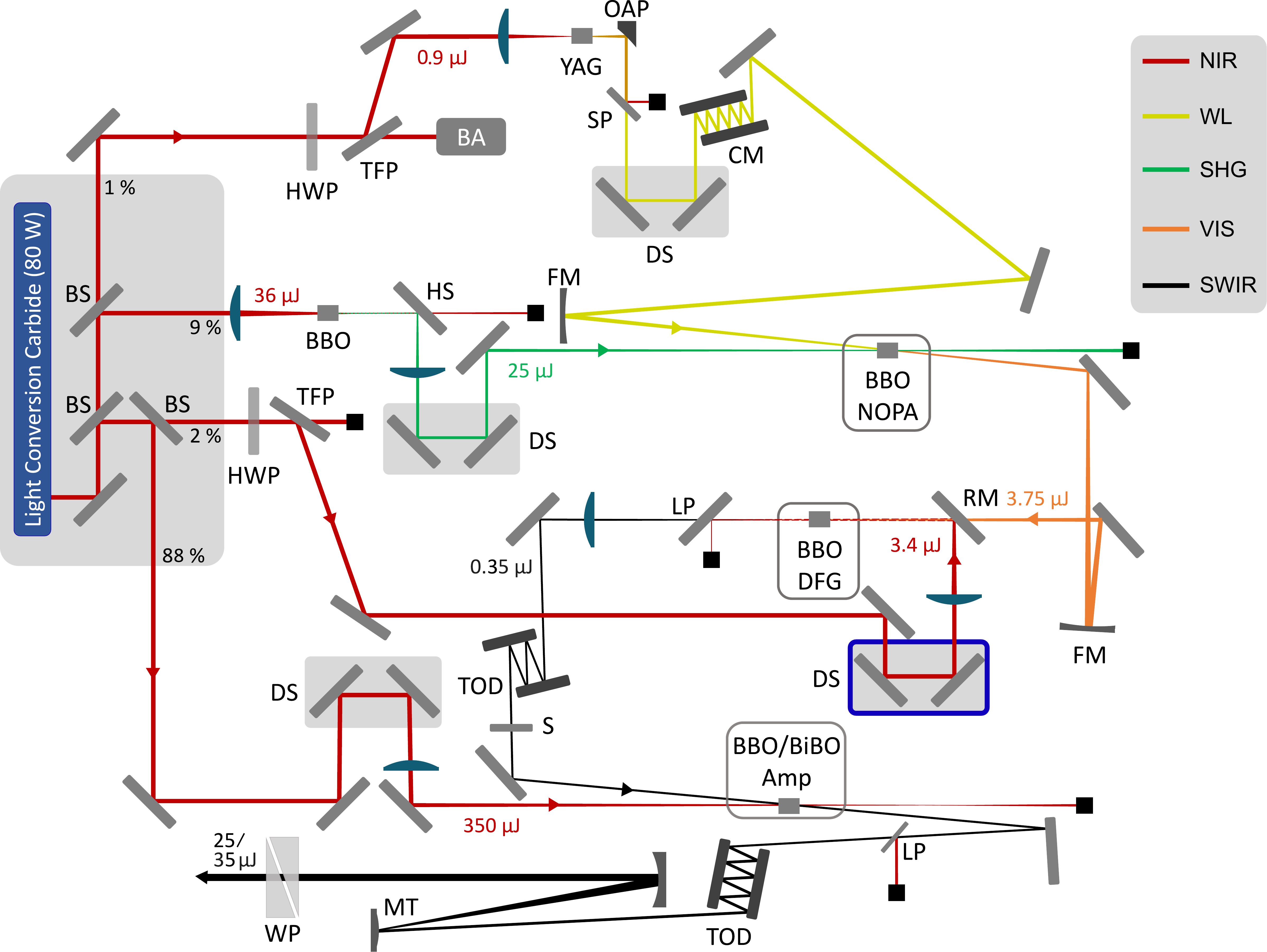}
\caption{\label{fig:setup} Detailed sketch of the laser system consisting of three main frequency conversion/amplification stages: NOPA:\:visible non-collinear optical parametric amplifier, DFG:\:difference frequency generation stage, Amp:\:optical parametric amplifier. The spectral ranges are encoded as: NIR: 1030 nm, WL: 450-780 nm, SHG: 515 nm, VIS: 625-740 nm (visible NOPA signal), SWIR: 1600-2500 nm. Additionally, the following abbreviations are used: BS:\:beamsplitter, HWP:\:half-wave plate, TFP:\:thin film polarizer, BA:\:beam alignment detector, YAG:\:yttrium aluminum garnet crystal, OAP:\:90° off-axis parabolic mirror, SP:\:short pass filter, LP:\:long pass filter, CM:\:pair of chirped mirrors, BBO:\:beta barium borate crystal, BiBO:\:bismuth triborate crystal, HS:\:harmonic separator, DS:\:delay stage, FM:\:focusing mirror, RM:\:recombining mirror, TOD:\:pair of third-order dispersive mirrors, S:\:sapphire plate (3\:mm), MT:\:mirror telescope, WP:\:wedge pair. The delay stage used for active CEP-stabilization is marked in blue.}
\end{figure*}

The system generating the 2-µm pulses is optically pumped by an 80-W Light Conversion Carbide laser, which is an ytterbium-based turnkey laser system consisting of a mode-locked oscillator, a regenerative amplifier and a stretcher-compressor unit. The pump laser pulses are characterized by a duration of 200\:fs, a central wavelength of 1030\:nm and a pulse energy of 400\:µJ at a repetition rate of 200\:kHz. Excellent pulse energy stability with a variation of 0.04\:\% measured in single-shot over 1\:min was found for these near-infrared (NIR) pulses. A detailed sketch of the complete laser system consisting of the pump laser and the frequency conversion and amplification stages is shown in Fig.\:\ref{fig:setup}. Starting on the left, the pump beam is divided into several parts in order to use them for the different frequency conversion stages. We separate a small part of the 1030-nm pump light from the rest using two 90:10 beamsplitters (BS) and a subsequent combination of a half-wave plate (HWP) and a thin film polarizer (TFP). The resulting pulses with an energy of just 1.4\:µJ are sufficient to generate a supercontinuum by focusing them into a 4\:mm long yttrium aluminum garnet (YAG) crystal. For supercontinuum generation, a lens with a focal length of 100\:mm is used, creating a filament in the YAG crystal. The white light (WL) is collimated by a 90-degree off-axis parabolic mirror (OAP) with an effective focal length of 12.7\:mm. \par
We frequency-double another part of the remaining 1030-nm light (about 36\:µJ) via a beta barium borate (BBO) crystal and use it as the pump beam in a visible non-collinear optical parametric amplifier (NOPA). By implementing the Poynting-vector compensated geometry with a non-collinear angle $\alpha$ of about 1.7$^\circ$ between the pump and the seed beams internally, it is possible to amplify a broad spectral range of the white light seed \cite{Cerullo_2003, Wilhelm_1997, Cerullo_1997}. As the nonlinear medium for amplification, we choose a BBO crystal with a thickness of 4\:mm cut at $\theta=23.6^\circ$ (type-I phase matching). The WL dispersion is adjusted so that its pulse duration is comparable to the pump pulse duration, resulting in five bounces on a pair of chirped mirrors (CM, Venteon DCM7). We use full width at half maximum (FWHM) focus sizes in the crystal of 620\:µm and 390\:µm for the pump and seed beam, respectively, which corresponds to a pump peak intensity in the BBO crystal of about 50\:GW/cm$^2$. As the focusing optics, a concave mirror with a radius of curvature of -800\:mm for the white light and a refocusing lens with a focal length of 350\:mm for the green pump beam are used.\par 
The NOPA output pulses, which are in the visible spectral range and have a pulse energy of 3.75\:µJ, are used in a subsequent difference frequency generation (DFG) stage. In this stage, a portion of the 1030-nm pump laser light (about 3.4\:µJ) is used as a seed to generate pulses in the short-wave infrared (SWIR) wavelength range. We focus the seed beam with a lens with a focal length of 500\:mm into a BBO crystal with a thickness of 1\:mm and a cutting angle of 20.4$^\circ$ (type-I phase matching). Similarly, we refocus the NOPA signal approximately 1\:m behind the NOPA crystal using a concave mirror with a radius of curvature of \mbox{-600\:mm}. A recombination mirror (RM, Edmund Optics \#14-004) is then used to collinearly overlap the 1030-nm seed and NOPA signal beams. The seed pulse energy as well as the FWHM spot sizes of the pump and the seed beams are adjusted so that the highest output pulse energy is reached while maintaining a good beam profile. This results in FWHM spot sizes of 515\:µm for the pump and 475\:µm for the seed, corresponding to peak intensities of 35\:GW/cm$^2$ and 18\:GW/cm$^2$, respectively. The resulting output pulses are in the SWIR spectral range centered around 2\:µm and have a pulse energy of 0.35\:µJ. \par
In the final nonlinear interaction stage, we amplify the idler pulses from the DFG stage to achieve the high pulse energy required for strong-field experiments in gases. The remaining 350\:µJ of the pump laser pulses are used for pumping this stage. Before amplification, the dispersion of the idler pulses is adjusted using a pair of third-order dispersive (TOD) mirrors (Ultrafast Innovations TOD2102, GDD/bounce: -100\:fs$^{2}$, TOD/bounce: -2000\:fs$^{3}$ at 2\:µm) and a sapphire plate with a thickness of 3\:mm (S). We do not optimize the idler dispersion for the shortest pulse duration before amplification, but for the best amplification in terms of spectral width. The amplification is realized with an optical-parametric amplifier (OPA) with a small non-collinear angle of 0.6$^\circ$ externally between the pump and the seed, which helps to separate them afterwards. For this stage, a BBO crystal cut at $\theta=21.6^\circ$ or a bismuth triborate (BiBO) crystal cut at $\theta=7.5^\circ$ and $\phi=0^\circ$ can be used as the nonlinear medium, both with a thickness of 4\:mm. The FWHM spot sizes of the pump beam, i.e. the 1030-nm pulses, and the seed beam, i.e. the DFG idler, in the crystal are 1780\:µm and 1640\:µm, respectively. They are achieved with a 1000-mm focal length lens for the pump and a 200-mm focal length CaF\textsubscript{2}-lens for the seed, which provides low dispersion at 2\:µm and enables collimation and only loose refocusing of the DFG idler into the amplification stage. The beam waist of the 1030-nm pump beam is a few centimeters behind the amplifier crystal. The pump peak intensity in the crystal reached with this configuration is approximately 70\:GW/cm$^2$. Finally, a mirror telescope is set up in order to perform a low dispersion beam size adjustment for the following experimental stages. The dispersion of the output pulses can be controlled with the help of another pair of TOD mirrors and a ZnS wedge pair.

\section{Results and Discussion}\label{sec2}
\subsection{Generation of SWIR pulses}

The white light generation in the YAG crystal delivers a broadband supercontinuum which reaches down to 530\:nm. It has a pulse energy of about 3\:nJ in the spectral range of 450-780\:nm and an average RMS pulse energy stability of 0.068($\pm0.005$)\%, measured for 2,000 consecutive white light pulses and repeated for a total of 250 times.\par
Part of the white light spectrum (625-740\:nm) is amplified in the first NOPA stage to a pulse energy of 3.75\:µJ, thus reaching an amplification factor of more than 1,000. The spectrum supports a transform-limited pulse duration of 12.3\:fs. Here and in the following, all pulse durations are defined as the FWHM width of the temporal intensity. The average RMS value for the pulse energy stability (20,000 pulses, repeated 379 times) is 0.109($\pm0.005$)\%. Due to the parabolic shape of the phase matching curve, a spectrum also consisting of two pronounced peaks and a dip in the middle is obtained. To investigate the pulse duration of the visible output pulses of the NOPA, we use an interferometric frequency-resolved autocorrelation (IFRAC) measurement \cite{Stibenz_2005}, which can be found in the Supporting Information in Fig.\:S1. The IFRAC measurement of the NOPA signal reveals a temporal duration of 39\:fs. Optimal compression of the NOPA signal pulses is not necessary at this stage because the seed pulses in the next conversion stage are much longer than 39\:fs anyway.\par
\begin{figure} [H]
\centering
\includegraphics[width=1\linewidth]{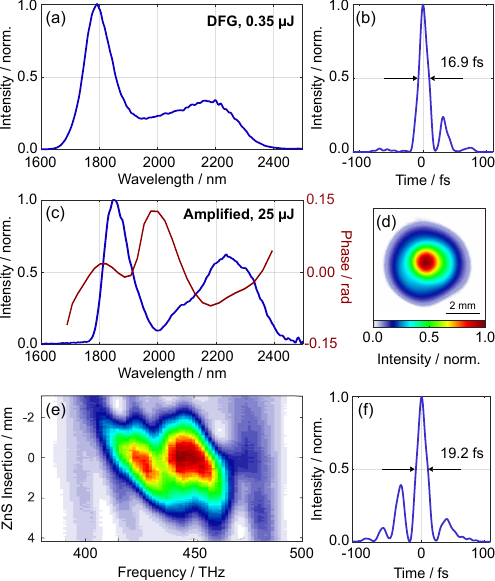}
\caption{\label{fig:2} Characterization of the SWIR pulses. (a) Measured spectrum of the DFG idler pulses, which have a pulse energy of 0.35\:µJ. (b) Retrieved temporal shape of the compressed DFG pulses (from a THG d-scan), revealing a pulse duration of 16.9\:fs. (c) Measured spectrum of the amplified pulses (blue), which have a pulse energy of 25\:µJ. The spectral phase retrieved from a THG d-scan is shown in red. (d) Beam profile of the amplified pulses after $\sim$2\:m of propagation behind the BBO. (e) THG d-scan trace for the amplified pulses, revealing the temporal pulse shape shown in (f) with a pulse duration of 19.2\:fs.}
\end{figure}

In the next step, we generate pulses in the SWIR spectral range in a collinear DFG process. The resulting spectrum ranges from 1650-2400\:nm with a central wavelength of 2025\:nm (see Fig.\:\ref{fig:2}(a)) and supports a transform-limited pulse duration of 16.7\:fs. In our setup, we reach a maximum output pulse energy of 0.35\:µJ for the DFG idler. The average RMS value for the pulse energy stability (2,000 pulses, repeated 401 times) is 0.145($\pm0.006$)\%. To investigate the pulse duration of the DFG idler pulses, we use the dispersion-scan (d-scan) technique \cite{dscan_Lund, Sytcevich_2021} in combination with the common pulse retrieval algorithm (COPRA) \cite{Geib_2019}. 

 The nonlinear signal for the d-scan is generated by focusing the pulses onto the surface of a \mbox{0.2-mm-thin} fused silica plate, leading to third-order harmonic generation (THG) at the air-glass interface \cite{Tsang.1995}. The temporal characterization of the DFG idler before the amplifier crystal, measured via THG d-scan, reveals a double pulse structure with a duration of the main pulse of 45\:fs. The retrieved temporal pulse form is shown in the Supporting Information in section S.2, as well as the impact of the amplifier crystal material dispersion on it. As an optional beam path, the DFG idler pulses are further compressed and used without propagation through the amplifier crystal. The resulting pulse structure with a normalized RMS trace error for the retrieval of 1.2\:\% is shown in Fig.\:\ref{fig:2}(b). The pulse consists of a short main pulse and a lower-intensity post-pulse. The main pulse has a duration of 16.9\:fs, which is very close to the transform-limited pulse duration of 16.7\:fs and corresponds to only 2.5 cycles of the optical field at a central wavelength of 2\:µm.\par

\subsection{Amplification of SWIR pulses}

In the final stage of the laser system presented, we amplify the SWIR pulses in an optical-parametric amplifier. The nonlinear medium used is BBO in a type I phase matching geometry. In this configuration, amplification by a factor of more than 70 is possible. Due to the different gain curve in the OPA stage, the amplified spectrum is slightly shifted in comparison to the DFG output spectrum (Fig.\:\ref{fig:2}(c)). The beam profile after amplification shows no clear distortions after $\sim$2\:m of propagation behind the BBO crystal (Fig.\:\ref{fig:2}(d)).\par 
With the amplification stage, we reach output pulse energies of up to 25\:µJ, which corresponds to a power of 5\:W at 200\:kHz. The average RMS value for the pulse energy stability (2,000 pulses, repeated 393 times) is 0.19($\pm0.01$)\% (see also Table~\ref{tab1}). Also here, due to the broadband phase matching around the degeneracy point at $\sim$2\:µm, a spectrum with two pronounced peaks is obtained. This spectrum remains broad, spanning from 1750-2450\:nm with a central wavelength of 2100\:nm, and supports a transform-limited pulse duration of 18.1\:fs. Nevertheless, the dip between the two peaks is more prominent for the amplified pulses due to the amplifier's gain curve, which has a minimum at 2 µm. This consequently changes the spectral form further. We also investigate the actual pulse duration of the amplified signal pulses using THG d-scan. The measured trace is shown in Fig.\:\ref{fig:2}(e). The retrieval of the data has an RMS trace error of 1.2\:\%. The retrieved spectral phase is shown in Fig.\:\ref{fig:2}(c) and the resulting temporal pulse structure in Fig.\:\ref{fig:2}(f). A short main pulse as well as lower-intensity side pulses are visible, which are correlated to the dip in the middle of the spectrum. This effect is further illustrated in the Supporting Information in section S.4. The additional asymmetry in the time structure is due to third-order dispersion. The TOD of the amplified pulses and of the (optionally non-amplified) DFG pulses is compensated using the same optics. As a compromise, the TOD is not perfectly compensated in neither case, but the flexibility of the system for different applications is greatly enhanced. The retrieved pulse duration of the main pulse is 19.2\:fs, which is close to the Fourier-limit and corresponds to less than 3 cycles of the optical field at a central wavelength of 2\:µm. \par
We further replaced the BBO crystal in the final amplification stage by a BiBO crystal as suggested by others \cite{NeuhausLaser}. In this case, we reach pulse energies of up to 35\:µJ with a comparable broad output spectrum. Additionally, the compressed main pulse has a slightly shorter retrieved duration of 18.1\:fs in comparison to BBO. Lower-intensity side-pulses remain prominent in this case as well. The measurements for characterizing the pulses after amplification in BiBO can be found in the Supporting Information in section S.3. In contrast to amplification in BBO, the thermal stabilization process in the BiBO crystal takes a few minutes. This means that the phase matching angle must be adjusted again afterwards. We attribute this to the crystal heating up as soon as the amplification process begins, which is in accordance with observations reported in \cite{NeuhausLaser}. Additionally, we observe less stable operation in terms of output power during the first few hours of use, which is probably due to an ongoing thermalization process resulting in higher temperature sensitivity of the system. For this reason, the rest of this work focuses on SWIR pulses amplified in BBO.

\begin{table}[h]
\caption{Pulse energy stabilities of the output pulses of the different stages.}\label{tab1}%
\begin{tabular}{@{}llll@{}}
\toprule
Stage & Averaged RMS & Variation of the RMS\\
\midrule
White light & 0.068\:\% & $\pm0.005$\:\%\\
NOPA & 0.109\:\% & $\pm0.005$\:\% \\
DFG & 0.145\:\% & $\pm0.006$\:\%\\
Amplifier & 0.190\:\% & $\pm0.010$\:\% \\
\botrule
\end{tabular}
\end{table}

\subsection{Carrier-envelope phase stability}

\begin{figure*}
\includegraphics[width=1\linewidth]{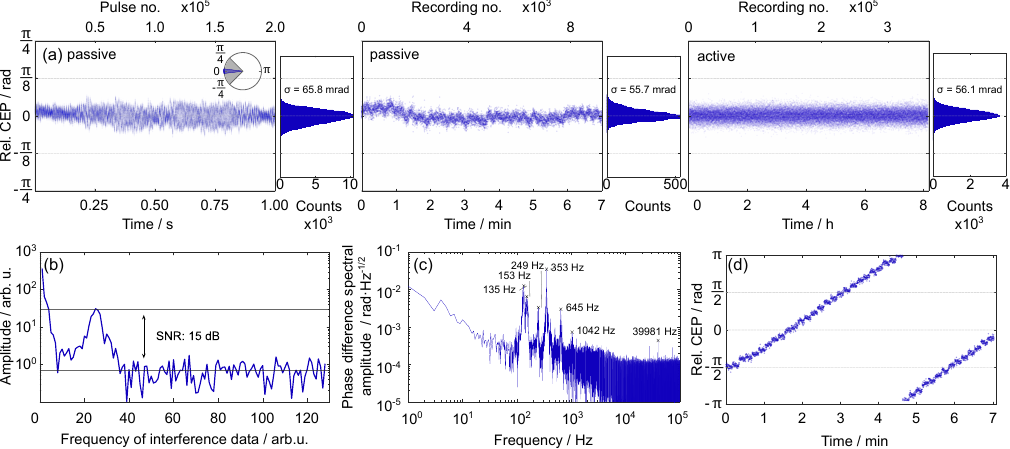}
\caption{\label{fig3} Characterization of the carrier-envelope phase of the amplified pulses. (a) Left: Single-shot CEP measurement, measured over 1\:s (i.e. of 200,000 pulses). The RMS CEP noise value is 65.8\:mrad. The inset illustrates the CEP values in a polar plot. Center: Passive CEP stability measurement, measured over 7\:min and averaging over 200\:pulses each. The RMS CEP noise value is 55.7\:mrad. Right: Long-term CEP measurement with active CEP stabilization, revealing an RMS noise value of 56.1\:mrad (averaged over 200\:pulses) over more than 8\:hours. (b) One example Fourier-transformed $f$-2$f$-spectrum. The peak, at which the phase is extracted, has a difference in the amplitude of 15\:dB to the background noise. (c) Complete noise spectrum of the measured CEP values shown in a) on the left. Some prominent peaks are marked with a black cross and their frequencies are given. (d) Stepwise CEP variation via a delay stage located in the seed arm of the DFG setup demonstrating the possibility of choosing the relative CEP value for experiments.}
\end{figure*}

As discussed above, strong-field charge carrier dynamics in gases and condensed matter is highly sensitive to the waveform of the electric field, which must be stable to observe the field-driven effects in temporally integrating measurements. To investigate the carrier-envelope phase stability of the presented laser system, we use a home-built $f$-$2f$-interferometer. It consists of a white light generation part and subsequent frequency doubling of the broadened spectrum, both producing spectral components in the region around 700-800\:nm. The interference of these two contributions behind a polarizer introduces CEP-dependent fringes in the spectrum. The presented laser system is passively CEP stable due to the specific DFG process used to generate the SWIR pulses, in which the two pulses share the same pump laser CEP changes \cite{Baltuska.2002}.\newline
We investigate the passive CEP stability of the output pulses after amplification in BBO in detail on different time scales. The results are shown in Fig.\:\ref{fig3}. We perform a short-term stability measurement on a shot-to-shot base with a fast line camera and the results are shown in Fig.\:\ref{fig3}(a) on the left. From the recorded fringes, the CEP of the SWIR pulses can be deduced via performing an inverse Fourier transform and extracting the phase at the fringe oscillation frequency, i.e. at the delay introduced by the optics in the f-2f-interferometer. The short-term CEP stability, measured over a period of 1\:s for 200,000 consecutive pulses, has an RMS noise value of 65.8\:mrad. An example Fourier transform of one $f$-$2f$-spectrum is shown in Fig.\:\ref{fig3}(b). From this, a signal-to-noise ratio (SNR) of about 15\:dB is determined. The measured CEP noise value is comparable to the highest stability level of an amplified system described by Thiré \textit{et al.}, where the authors report on a 3.2-µm amplifier with 40\:fs pulses and a single pulse CEP stability of 65\:mrad \cite{Thire.2018}. In a more similar system described by Homann \textit{et al.}, a short-term CEP stability of the 10-fs, 1.8-µm pulses of 78.5\:mrad measured over 0.5\:s is reported \cite{Homann.2012}. This value is slightly higher than the value reported here, although in \cite{Homann.2012} 100 pulses of the 100-kHz pulse train have been averaged. Since we recorded the CEP for each individual shot, we can calculate a CEP stability averaging over an arbitrary number of pulses. When we add up 100 of our measured single-shot spectra each, thus reducing the number of spectra from 200,000 to 2,000, we reach an RMS value of 60.8 mrad for the 1-second measurement. This further improves the value we found for each individual shot, but at the same time reduces the validity concerning high frequency noise. \par 
Our single-shot CEP values show a small, but clear oscillation in their time evolution. We compared the CEP fluctuations to a simultaneous shot-to-shot measurement of the pulse energy and did not find a significant correlation. The measurement is shown in section S.6 of the Supporting Information. In order to investigate the oscillation frequencies further, a spectral analysis of the CEP noise is performed. It is shown in Fig.\:\ref{fig3}(c) and some prominent noise frequencies are marked with black crosses. There are low-frequency noise contributions at about  135\:Hz, 150\:Hz, 250\:Hz, 350\:Hz, 645\:Hz and 1040\:Hz, which correspond to acoustic noise in the lab.
Additionally, high-frequency noise components are measured, out of which the contribution at about 40\:kHz is most pronounced. Investigating noise contributions at such high frequencies is only possible due to the high laser repetition rate and the high acquisition rate of the shot-to-shot measurement in this work. Other works also characterize the CEP of single pulses, but not on a shot-by-shot basis. For example, in \cite{Thire.2018}, the authors detect every 10th pulse for a laser repetition rate of 100\:kHz. In this case, high-frequency noise can be undersampled and the CEP noise may be underestimated. This effect occurs for oscillation frequencies which are integer multiples of the sampling frequency, which can easily be the case for electronic noise which is common at round frequencies.\par
In a next step, we perform a CEP measurement on a longer timescale, again without any active stabilization of the CEP. With this method, also slow CEP drifts and noise contributions with a lower frequency can be determined compared to the previous 1-second measurement. At the same time, a high passive CEP stability on a minute scale is important to establish a reliable active stabilization scheme for CEP-dependent experiments on an hour to day scale. We record $f$-$2f$-spectra using a spectrometer with an integration time of 1\:ms for 7 minutes. With the given repetition rate of 200\:kHz, each spectrum contains averaged information about the CEP of 200 pulses. The measured spectra for the 7-minute measurement are presented in Fig.\:S6 of the Supporting Information. As before, an inverse Fourier transform is applied to the spectra to extract the phase of the oscillation for each spectrum.  The resulting relative CEP values are shown in Fig.\:\ref{fig3}(a) in the center. A slight drift of about 60\:mrad from the beginning to the end of the measurement can be observed, which is related to small changes in the lab environment such as air convection. The RMS value of the CEP noise is calculated from the extracted phases and is found to be 55.7\:mrad. Due to averaging over 200 pulses, the SNR for each measurement improves and fast CEP variations are averaged out. This results in a better RMS CEP noise value than in the shot-to-shot measurement despite the small drift. The noise value is lower than the RMS value for a minute-scale measurement reported for a comparable laser system, also with a central wavelength of 2\:µm and a repetition rate of 200\:kHz \cite{LundLaser}. The authors report an RMS value of 290\:mrad averaged over 300 pulses and measured over 10\:min.\par
Our second- and minute-scale measurements can be compared by averaging over 200 consecutive pulses of the single-shot dataset. When 200 of the retrieved CEP values are averaged each, we reach an RMS value of 54.2\:mrad for the 1-second measurement. A more realistic approach is to sum up 200 consecutive spectra, as intrinsically done by a spectrometer with a longer integration time, and then determine the resulting CEP value. In our case, this leads to a RMS value of 54.7\:mrad for the 1-second measurement. Both values compare well to the minute-long measurement value of 55.7\:mrad.\par

Due to intensity-to-phase coupling during the WLG in the seed arm of the NOPA, the carrier-envelope phase can change over time \cite{9191090}. This, along with other factors affecting the CEP in only one of the two arms of the DFG stage, results in a change to the CEP of the DFG idler pulses. It can be compensated with the help of a feedback-loop for a delay stage located in the IR arm of the DFG (marked in blue in Fig.\:\ref{fig:setup}), thus shifting the seed of the DFG relative to the pump. With this possibility of compensating the CEP drift over time, a high long-term CEP stability is reached, as shown in Fig.\:\ref{fig3}(a) on the right. There, the time evolution of the actively stabilized CEP over more than eight hours is shown, after which the measurement was concluded. The data was collected in the same way as in the 7-min measurement, so that still an averaging over 200 pulses is done. The RMS value for this measurement is 56.1\:mrad, demonstrating that excellent passive CEP stability can be maintained over an extended period of time.\par
The achieved long-term CEP stability is higher than the RMS value of 242\:mrad measured by Neuhaus \textit{et al.} for a period of 30\:min and of 250\:mrad measured by Ritzkowsky \textit{et al.} for a period of 60\:min \cite{NeuhausLaser, Ritzkowsky.2026}. Quite notably, Kurihara \textit{et al.} reported on a stability of 92\:mrad averaging over 1000 pulses over a period of 11 hours without any active stabilization, reaching only a slightly higher value than what is demonstrated in our work with an active feedback loop \cite{Kurihara.2023}.\par
Finally, Fig.\:\ref{fig3}(d) shows an incremental linear change in the carrier-envelope phase of more than $2\pi$. This demonstrates the feasibility of selecting the relative CEP of SWIR pulses for subsequent strong-field experiments. In summary, the feedback loop for active CEP stabilization allows experiments to be performed over several hours with high CEP stability below 60\:mrad at a chosen relative CEP value.\par

\section{Application: HHG in argon}

Due to the high pulse energy at long wavelengths, the pulses generated at a high repetition rate are ideally suited for strong-field experiments. An example is high-order harmonic generation of the driving wavelength in gases, where the pulses are used to generate a broad spectrum in the XUV spectral range. We focus the 2-µm pulses tightly into an argon gas jet from a home-built gas target with a gas nozzle opening of 50\:µm and a backing pressure of 15\:bar. The light emerging from the interaction region is directed to a spectrometer consisting of a concave aberration-corrected flat-field grating (Hitachi 001-0437), a microchannel plate and a phosphor screen.\par
\begin{figure} [H]
\centering
\includegraphics[width=1.0\linewidth]{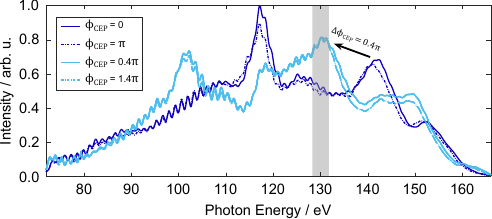}
\caption{\label{fig:hhg} High-order harmonic spectra recorded for four different values of the relative carrier-envelope phase of the driving pulses. No absolute CEP values are determined, but an arbitrary CEP value is set to $\upphi_{CEP} = 0$. The CEP-dependent shape of the spectrum at the cutoff region shows a $\uppi$-periodicity. The vertical line at 130\:eV represents an experimentally desired spectral range in our laboratory. The CEP can be chosen so that maximum intensity in the desired spectral range is achieved.}
\end{figure}

Harmonic spectra for four different values of the CEP of the driving pulses are recorded and shown in Fig.\:\ref{fig:hhg}. As only the relative CEP can be measured with the $f$-$2f$-interferometer, no absolute CEP value is determined. For illustration, we set an arbitrary CEP value to $\upphi_{CEP} = 0$. As can be seen in the figure, a high cut-off photon energy of up to 160\:eV is reached, which is limited by the driving laser intensity used in the experiment. The high-order harmonic generation process is highly sensitive to the relative value of the CEP of the driving pulses, particularly in the cut-off region. Shifting the CEP changes the position of the spectral maxima to different photon energies. All spectra are normalized to the maximum intensity of the spectrum with $\upphi_{CEP} = 0$, which means that the signal intensities (and thus the corresponding XUV flux) is directly comparable. The CEP-dependent peaks exhibit a $\uppi$-periodicity, as illustrated in the figure for the two relative CEP values of 0/$\uppi$ and 0.4$\uppi$/1.4$\uppi$. This effect can be explained by high-order harmonic emission from individual half-cycles of the driving laser pulse, as discussed in the literature \cite{Gaumnitz_2017,Haworth.2007}. Notably, in our setup, the half-cycle cutoff (HCO) effects are clearly visible and, in contrast to \cite{Haworth.2007}, also prominent without active CEP stabilization due to the high passive CEP stability of the SWIR pulses.\par

For an application in photoemission electron microscopy (PEEM), an optical setup for a central photon energy of 130 eV was designed. This spectral region is marked in Fig.\:\ref{fig:hhg}. While the integrated flux over the entire spectrum is just 2\:\% higher for a relative CEP value of $0.4\uppi$~($1.4\uppi$) in comparison to a value of $0$~($\uppi$), the flux in the marked spectral region is increased by 60\:\%. Changing the relative CEP from $0$~($\uppi$) to $0.4\uppi$~($1.4\uppi$) shifts the HCO peak at more than 140\:eV to the desired spectral range, as indicated by the black arrow. The CEP dependency of the HHG spectra can be used to maximize the signal in a specific spectral region. Also, as shown in \cite{Haworth.2007}, an extraction of the absolute CEP of the generating laser pulse from the measured XUV spectra is possible with some assumptions.\par
The photon flux of the XUV light is estimated from the measured data by taking into account the efficiencies of the components used. It is approximately $5\cdot10^7$ photons/s in the spectral region of 70-180\:eV. This corresponds to an HHG conversion efficiency to this part of the spectrum of $4\cdot10^{-10}$. The rest of the spectrum below 70\:eV is not recorded with our spectrometer setup, which is why it is neglected for the efficiency calculations. This means that the calculated efficiency is only a lower limit for the conversion efficiency into the whole XUV spectrum. If more XUV flux is required, the generation parameters could be further optimized following Weissenbilder \textit{et al.} \cite{Weissenbilder.2022}. Increasing the pressure in the argon jet should improve the conversion efficiency for our focusing geometry. Until now, a turbo pump with a pumping speed of 300\:l/s has been used in the HHG chamber. In the near future, adding a turbo pump with a higher pumping speed to this setup will make it possible to reach a higher argon pressure in the interaction regime.\par
As surface science experiments benefit from high repetition rates to avoid space charge effects and provide rather high interaction cross sections, already the moderate XUV pulse energies generated here are suitable for, for example, time-resolved XUV-IR PEEM-experiments on 2D materials and nanostructures. This enables the investigation of ultrafast charge carrier dynamics across nanoscale interfaces of metals and semiconductors \cite{Vogelsang_2024, Vogelsang.2025}.

\section{Conclusion}

In summary, we present a conceptually simple and compact laser system generating few-cycle laser pulses at a central wavelength of 2\:µm. Pumped by an ytterbium-based system with 200-fs pulses at a central wavelength of 1030\:nm, we produce sub-20-fs pulses with a pulse energy of 25\:µJ at a repetition rate of 200\:kHz. The output pulses exhibit excellent passive CEP stability with an RMS CEP noise value characterized in depth on time scales ranging from microseconds to hours.

A 1-second measurement resolves the CEP for each pulse of the 200-kHz pulse train and provides an excellent CEP stability of 65.8\:mrad. It demonstrates the low amplitude of high-frequency noise in our laser system. This high-frequency contribution could be overlooked in slower measurements averaging over multiple pulses, even though it may have a significant impact on CEP stability. Additionally, we did not find a correlation of the measured CEP values with the pulse energy. We conclude that amplitude noise of the laser is expected to have a low impact on the $f$-$2f$-measurement and thus, the measured RMS value is presumed to represent an upper bound, limited by the signal-to-noise ratio of 15\:dB of the measurement, with the true CEP stability possibly being even higher. A 7-minute measurement uses averaging to keep the data rate at an acceptable level and probes potential CEP changes on a time scale relevant for an active feedback loop further stabilizing the CEP. On this scale, we find a high stability of 55.7\:mrad, demonstrating that an active feedback loop can handle the remaining CEP drifts. In a final 8-hour measurement, we demonstrate long term stability using an active feedback loop, proving the readiness of the system for experiments with high demands on statistics.

Photoemission from gases and condensed matter is a large group of experiments that require long measurement times and, consequently, a stable driving system - in particular regarding the electric field waveform. We demonstrate this by showing distinctly different HHG spectra for varying CEP values, which exhibit a much richer structure than the often-observed featureless plateau region with a cutoff at 3.17 times the ponderomotive energy. The high CEP stability enables this observation without active stabilization and makes a detailed investigation of the HHG process possible. In the future, more sophisticated control of stable light waves will most likely permit indirect means to alter high-order harmonic spectra up to the x-ray spectral range, where only very basic optical tools are available.

\backmatter

\bmhead{Supplementary information}

Supplementary material is available.

\bmhead{Author contribution}

The work was conceptualized by J.V. The laser system was set up and characterized by K.M., A.K., K.H., L.H. and L.B. C.L.A. and C.G. have developed the single-shot CEP detector. The single-shot CEP measurements were performed by K.M., C.G., C.J., J.N. and J.V. The initial draft of the manuscript was prepared by K.M. and J.V., and all authors contributed to the final version.

\bmhead{Acknowledgements}

The authors would like to thank TEM Messtechnik for the excellent Aligna beam pointing stabilization system and the Attosecond Science group at Lund University for the TOD mirrors.
Furthermore, the authors would like to thank Gunnar Arisholm for the SISYPHOS simulation code and fruitful discussions.

\bmhead{Funding}

This study was funded by the zukunft.niedersachsen program of the Niedersächsisches Ministerium für Wissenschaft und Kultur (DyNano and Stay Inspired) and the Deutsche Forschungsgemeinschaft (462448709, Emmy Noether program).

\bmhead{Data availability} 

Data underlying the results presented in this paper are not publicly available at this time but may be obtained from the authors upon reasonable request.

\section*{Declarations}

\bmhead{Conflict of interest/Competing interests}
The authors declare no competing interests.

\bmhead{Ethics approval and consent to participate}
Not applicable.

\bmhead{Consent for publication}
Not applicable.

\bmhead{Materials availability}
Not applicable.

\bmhead{Code availability}
Not applicable.

\bibliography{Laserpaper}

\newpage

\onecolumn
\newgeometry{
left = 20mm,
right = 20mm,
top = 20mm}
\renewcommand{\thesection}{S}
\pagenumbering{gobble}

\Large
\begin{center}
Supplementary Material:\\ Multiscale carrier-envelope phase characterization of 2-µm pulses\\ delivered by a 200-kHz optical parametric amplifier\\

\hspace{5pt}

\large
Katrin Meier$^1$, Arvid Klösgen$^1$, Kerstin Harland$^1$, Lina Hansen$^1$, Lars Barnekow$^1$,\\ Chen Guo$^2$, Caroline Juliano$^2$, Jin Niu$^2$, Cord L. Arnold$^2$ and Jan Vogelsang$^1$

\hspace{5pt}

\small  
$^1$Institute of Physics, Carl von Ossietzky Universität Oldenburg, Germany.\\
$^2$Department of Physics, Lund University, Sweden.\\

\end{center}
\normalsize

\noindent This document provides supplementary information to "Multiscale carrier-envelope phase characterization of 2-µm~pulses delivered by a 200-kHz optical parametric amplifier". Additional measurement data are shown to further characterize the signal pulses of the visible non-collinear optical parametic amplifier (NOPA), the idler pulses of the difference-frequency generation (DFG) stage and the amplified pulses when using BiBO as the nonlinear crystal. The temporal structure of the amplified near-infrared (NIR) pulses is further characterized by calculating the Fourier-transformation of the spectrum. Additionally, the procedure for measuring the carrier-envelope phase (CEP) of the output pulses is further explained.

\subsection{Temporal characterization of the VIS-NOPA}
The signal output pulses of the visible NOPA stage are characterized using an interferometric frequency-resolved autocorrelation (IFRAC) measurement \cite{Stibenz_2005}.

\begin{figure} [H]
    \centering
    \captionsetup{width=0.9\linewidth}
    \includegraphics[width=0.8\textwidth]{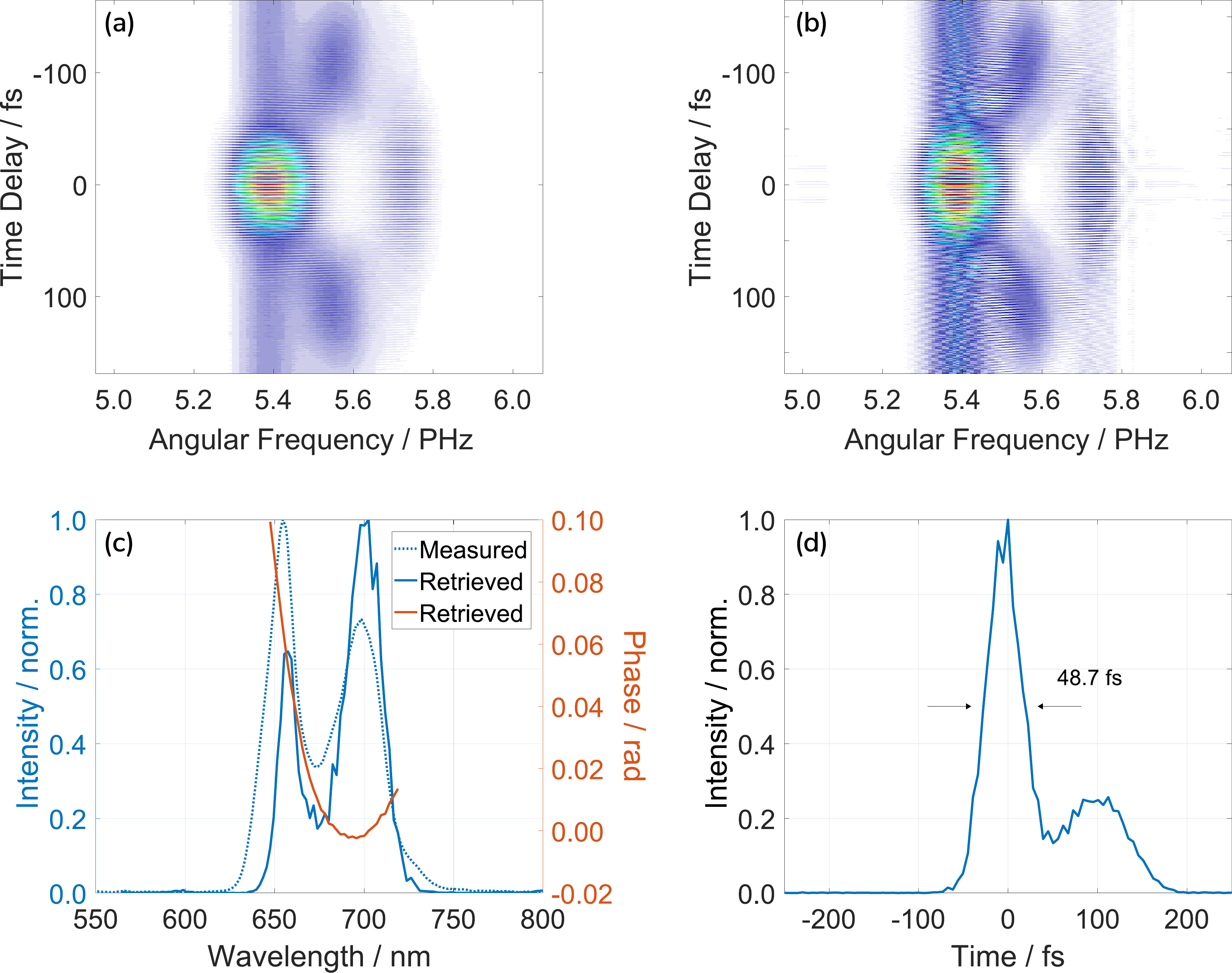}
    \caption*{Figure S1: Measured (a) and retrieved (b) IFRAC measurement for the characterization of the NOPA signal (retrieval trace error: 2.4\:\%) with the additional dispersion of the beamsplitter used in the interferometer (1\:mm of fused silica at an angle of 45°) included. The resulting spectral and temporal pulse shape are shown in (c) and (d), respectively.}
    \label{fig:nopa}
\end{figure}

\noindent The NOPA spectrum supports a transform-limited pulse duration of 12.3\:fs. The IFRAC trace is shown in Fig.\:S1(a) together with the retrieved trace (b) with a retrieval trace error of 2.4\:\%. The retrieved pulse duration is 48.7\:fs (d), which corresponds to a pulse duration of 39.0\:fs for the pulse itself after subtraction of the phase induced by 1 mm of fused silica at an angle of 45° used as a beamsplitter in the interferometer. The parabolic shape of the phase curve shown in Fig.\:S1(c) indicates a significant impact of group velocity dispersion $GDD$. This is further determined by fitting the phase data using a Taylor series around $\lambda_0= 680$\:nm revealing $|GDD| = 560$\:fs$^2$.

\subsection{Temporal characterization of the DFG idler pulses} 
The DFG idler pulses before the amplification stage are characterized via third-harmonic generation dispersion-scan (THG d-scan). The result is shown in Fig.\:S2(a). The main pulse has a FWHM pulse duration of 45\:fs, followed by a post-pulse with a relative height of 75\:\% and a FWHM pulse duration of 55\:fs. With the additional material dispersion of the amplifier crystal, the total pulse duration and the duration of each of the two sub-pulses decrease. This is also illustrated in Fig.\:S2(b). When propagating through the BBO crystal, the relative height of the post-pulse decreases to 70\:\% and the temporal FWHM of the main pulse decreases to 25\:fs. In the BiBO crystal, the relative height of the post-pulse decreases to 60\:\% and the temporal FWHM of the main pulse decreases to 25\:fs. 

\begin{figure} [h]
    \centering
    \captionsetup{width=1.0\linewidth}
    \includegraphics[width=1.0\linewidth]{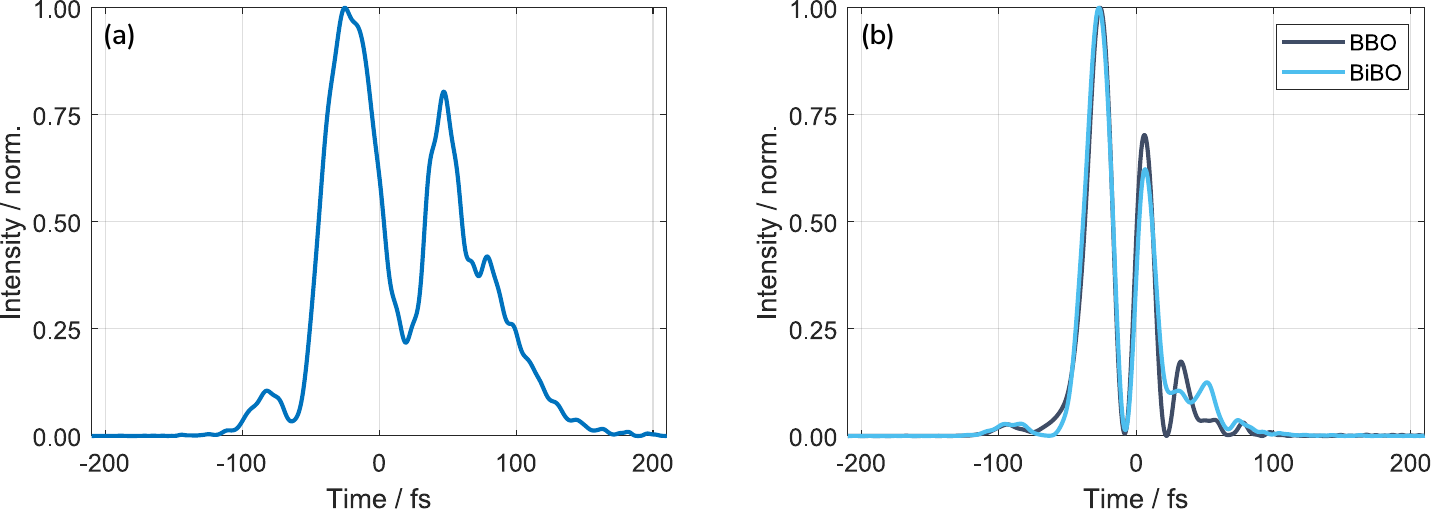}
    \caption*{Figure S2: Pulse duration of the DFG pulses (a) before the amplification stage, measured via THG d-scan, and (b) after calculated propagation through 4\:mm of BBO/BiBO.}
    \label{fig:dfg}
\end{figure}

\newpage

\subsection{Temporal characterization of the NIR pulses amplified in BiBO} 
The NIR pulses after amplification in BiBO are also characterized via THG d-scan. The result is shown in Fig.\:S3(a). The normalized RMS trace error for the retrieval is 1.5\:\%. As in the case of BBO, the dip in the middle of the spectrum is visible, as well as the correlated lower-intensity side pulses in the time domain, each with a peak intensity of about 25\:\% of the pulse peak intensity. The main pulse has a FWHM duration of 18.1\:fs, which is close to the transform-limited pulse duration of 16.7\:fs.

\begin{figure} [H]
    \centering
    \captionsetup{width=0.9\linewidth}
    \includegraphics[width=0.8\linewidth]{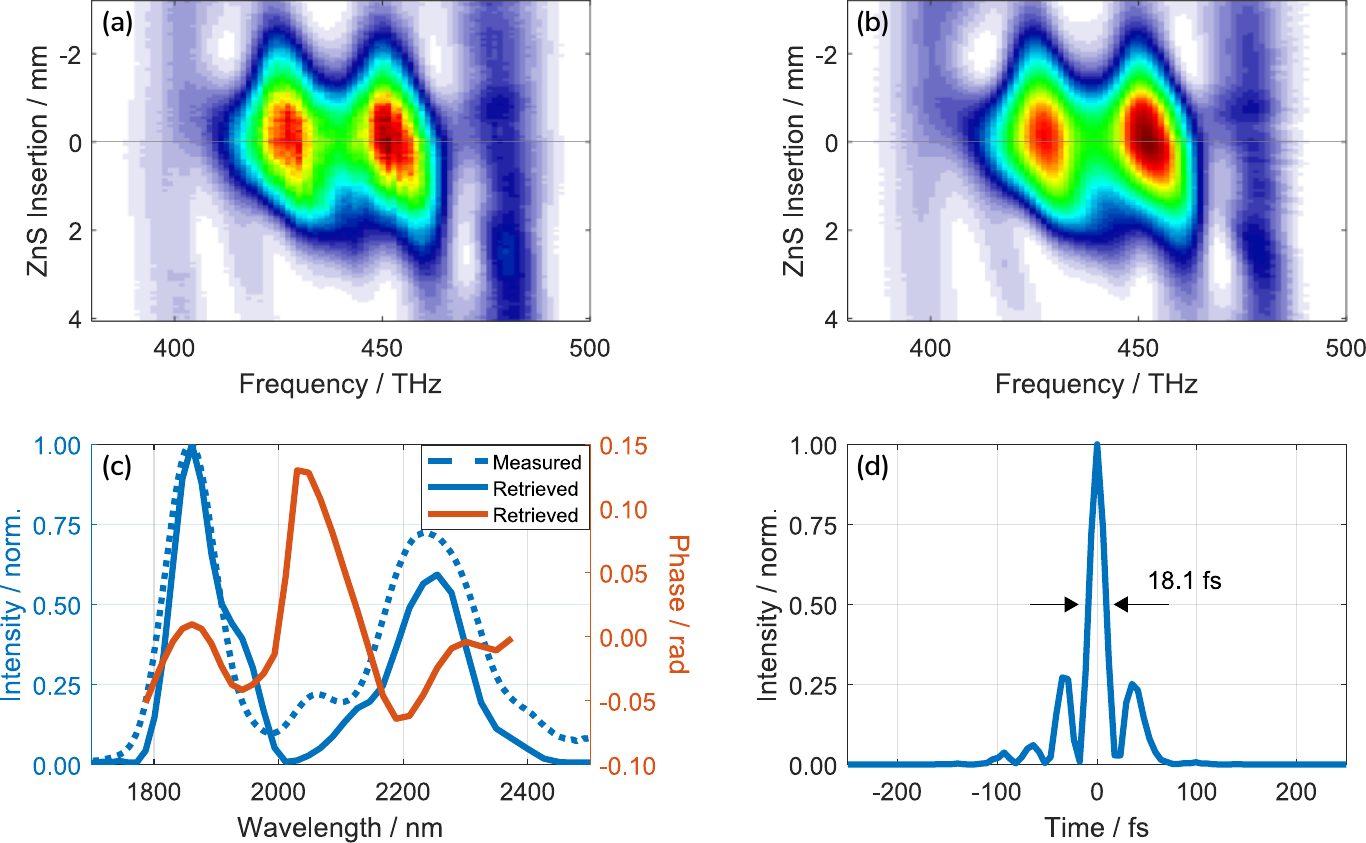}
    \caption*{Figure S3: (a) Measured and (b) retrieved d-scan trace of the NIR pulses amplified in BiBO (retrieval trace error: 1.5\:\%). (c) Measured fundamental spectrum and reconstructed spectral intensity and phase. (d) Resulting temporal structure, revealing a short main pulse with a temporal FWHM duration of 18.1\:fs, surrounded by lower-intensity side pulses.}
    \label{fig:bibo}
\end{figure}

\subsection{Transform-limited pulse duration of the amplified pulses}
The temporal shape of the amplified pulses exhibits symmetric, lower-intensity pre- and post-pulses. This relates to the shape of the fundamental spectrum, specifically the dip in the middle at around 2\;µm. This is illustrated by computing the temporal structure of a pulse with the measured spectral intensity and a flat spectral phase:

\begin{figure} [H]
    \centering
    \captionsetup{width=1.0\linewidth}
    \includegraphics[width=1.0\linewidth]{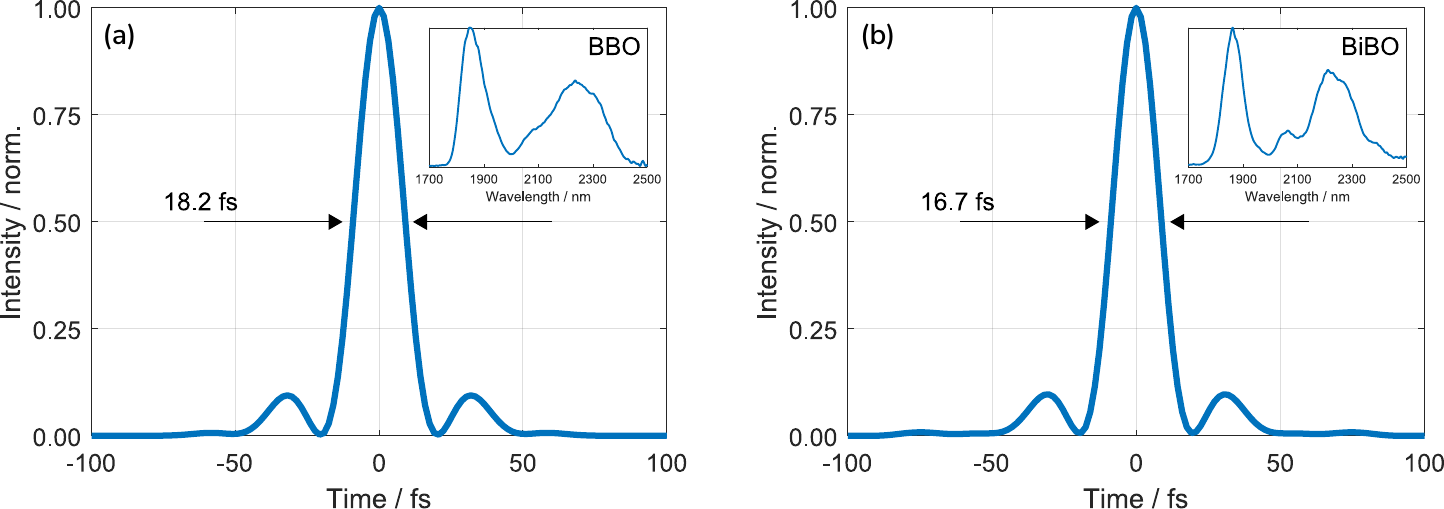}
    \caption*{Figure S4: Calculated temporal pulse structure for a laser pulse with the measured spectrum (inset) and a flat phase for (a) amplification in BBO and (b) in BiBO.}
    \label{fig:tlpulse}
\end{figure}

\subsection{Simulation of the amplification in BBO}
The parametric amplification process is simulated using the SISYFOS code package \cite{Sisyphos}. A complete experimental dataset including input and output pulse energies, spot sizes, spectra and temporal pulse structures is collected prior to running the simulation. The simulated output spectrum compared to the measured one is shown in Fig.\:S5. 

\begin{figure} [H]
    \centering
    \captionsetup{width=0.7\linewidth}
    \includegraphics[width=0.4\linewidth]{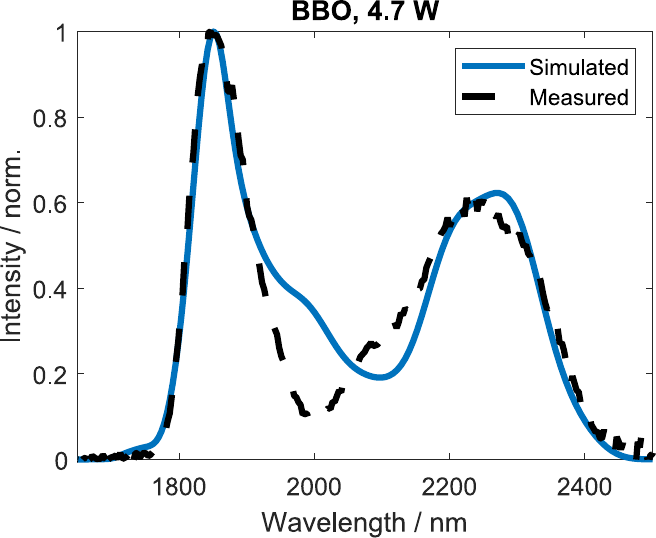}
    \caption*{Figure S5: Simulated output spectrum with a simulated output power of 4.7\:W in comparison to the measured output spectrum with a power of 4.4\:W after amplification in BBO.}
    \label{fig:tlpulse}
\end{figure}


\subsection{Investigation of the CEP stability of the laser system}
The CEP of the output pulses is measured using a home-built $f$-$2f$-interferometer. It consists of a white light generation part and subsequent frequency doubling of the broadened spectrum, both producing spectral components in the region around 700-800\:nm. The relative intensity of these components is adjusted using a polarizer. The resulting spectral interference pattern averaged over 200 pulses and recorded for seven minutes is shown in Fig.\:S6(a). An exemplary spectrum from the beginning of the measurement is shown in Fig.\:S6(b).

\begin{figure} [H]
    \centering
    \captionsetup{width=1.0\linewidth}
    \includegraphics[width=1.0\linewidth]{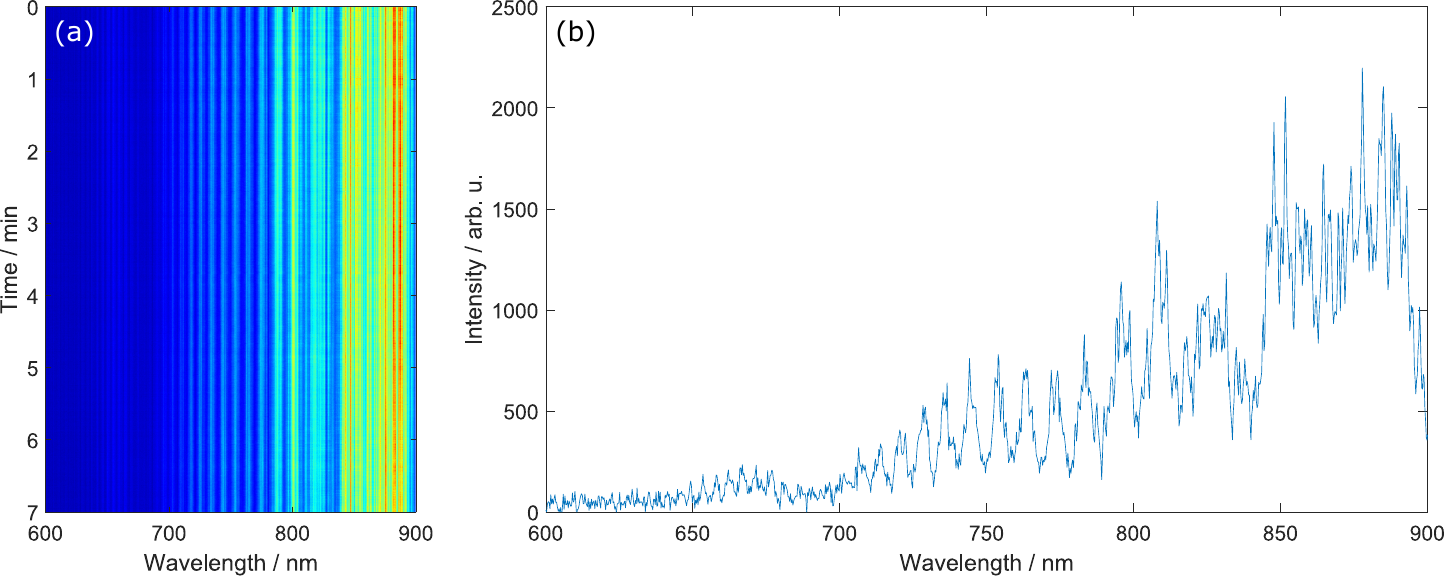}
    \caption*{Figure S6: Investigation of the CEP stability of the laser system. (a) Interference pattern recorded with an $f$-$2f$-interferometer averaged over 200 pulses and measured over seven minutes. (b) Exemplary spectrum from the dataset shown in (a). The phase of the fringe oscillation is extracted at the peak of the Fourier-transform of the signal in the spectral region of 700-800\:nm.}
    \label{fig:cep_passive}
\end{figure}

\noindent A possible correlation between the measured CEP oscillations and the pulse energy stability is investigated by measuring both simultaneously on a shot-to-shot basis. The result is shown in Fig.\:S7, which reveals no direct correlation between the two quantities. Additionally, a calculated cross-correlation revealed no evident connection. The recorded pulse energy stability over the measurement time frame is 1.6\:\%, which is higher than the value reported in the main manuscript. This discrepancy is due to suboptimal measurement conditions in a preliminary setup used here. Small beam pointing instabilities will most likely have influenced the pulse energy determination, which is why the dataset actually includes both pulse energy and beam pointing variations. No correlation is found, but a more detailed investigation with a separate measurement of all relevant parameters might be interesting to look at in the future. The impact of a shutter used to synchronize the measurements can be seen in the values for the final pulses.

\begin{figure} [H]
    \centering
    \captionsetup{width=1.0\linewidth}
    \includegraphics[width=1.0\linewidth]{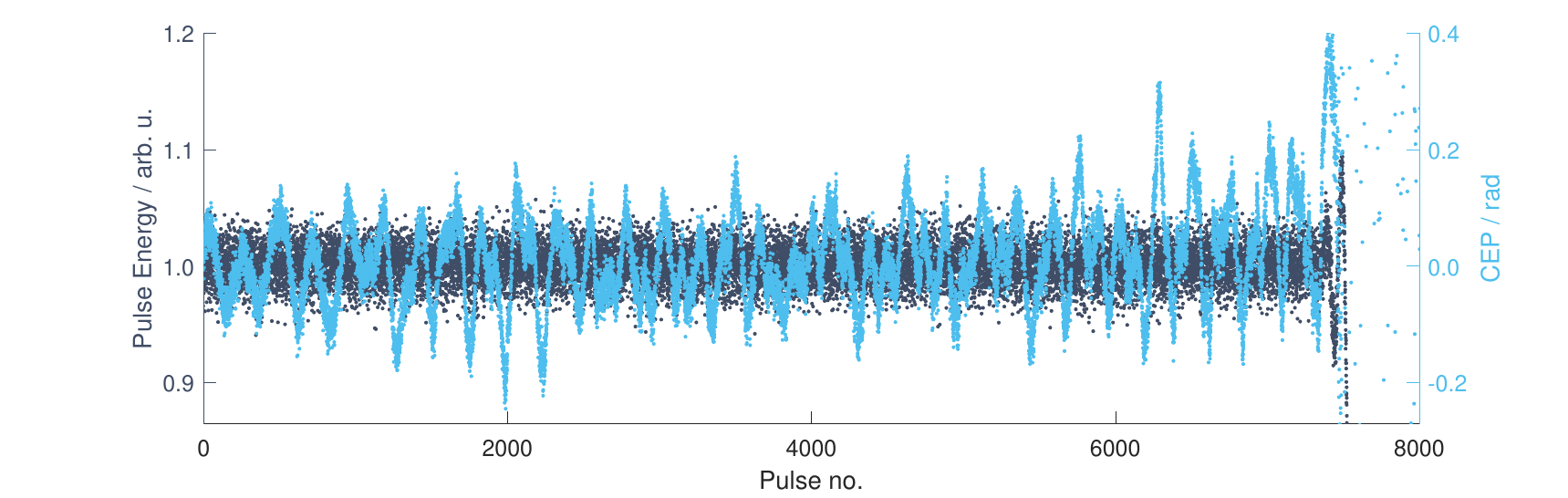}
    \caption*{Figure S7: Comparison of the CEP stability and the pulse energy stability.}
    \label{fig:cep_passive}
\end{figure}

\end{document}